\documentclass[12pt]{article}
\usepackage{amsmath}
\usepackage{amsfonts}

 \makeatletter
\@addtoreset{equation}{section}

\makeatother
\input{tcilatex}
\begin{document}

\bigskip \thispagestyle{empty}

\begin{center}
\null\vspace{-1cm} \hfill \\[0pt]
\vspace{1cm}{\large \ \ \ \textbf{SUPER GELFAND-DICKEY ALGEBRA \\[0pt]
AND INTEGRABLE MODELS}}

\bigskip {\textbf{A. El Boukili, M.B. Sedra, A. Zemate}}\\[0pt]

{\emph{Universit\'{e} Ibn Tofail, Facult\'{e} des Sciences, D\'{e}partement
de Physique,\\[0pt]
Laboratoire de Physique de la Mati\`{e}re et Rayonnement (LPMR), K\'{e}%
nitra, Morocco.}}\\[0pt]
\end{center}

We present in this work a systematic study of integrable models
and supersymmetric extensions of the Gelfand-Dickey algebra of
pseudo differential operators. We describe in detail the relation
existing between the algebra of super pseudo-differential
operators on the ring of superfields $u_{\frac{s}{2}}(z,\theta
),s\in Z$ and the higher and lower spin extensions of the
conformal algebra.

\section{Introduction}

$2d$-Integrable models \cite{1} in connection with conformal field theories
(CFT) \cite{2,3} and their underlying lower $(s\leq 2)$ \cite{4, 5, 6, 7, 8,
9} and higher $(s\geq 2)$ \cite{10,11} spin symmetries, have occupied a
central position in various areas of research. More particularly, a lot of
interest has been paid to $w$-symmetries \cite{10}, which are infinite
dimensional algebras extending the Virasoro algebra by adding to the energy
momentum operator $T(z)\equiv W_{2}$, a set of conserved currents $w_{s}(z)$%
, of conformal spin $s>2$ with some composite operators necessary for the
closure of the algebra.\newline
\newline
In the language of $2d$ CFT, the above mentioned currents $w_{s}$ are taken
in general as primary satisfying the OPE \cite{2}
\begin{equation}
T(z)W_{s}(\omega )=\frac{s}{(z-\omega )^{2}}W_{s}(\omega )+\frac{%
W_{s}^{\prime }(\omega )}{(z-\omega )},
\end{equation}%
or equivalently,
\begin{equation}
W_{s}=J^{s}{\tilde{W}}_{s}
\end{equation}%
under a general change of coordinate (diffeomorphism) $x\rightarrow {\tilde{x%
}}(x)$ with $J=\frac{\partial \tilde{x}}{\partial x}$ is the associated
Jacobian. These $w$-symmetries exhibit among others a non linear structure
and are not Lie algebra in the standard way as they incorporate composite
fields in their OPE. In integrable models these higher spin symmetries
appear such that the Virasoro algebra corresponds to the second Hamiltonian
structure (Gelfand Dickey Poisson Bracket) for the KdV hierarchy \cite{12,
13}, $w_{3}$ for the Boussinesq \cite{14} and $W_{1+\infty }$ for the KP
hierarchy \cite{15} and so on. These correspondences are achieved naturally
in terms of pseudo-differential Lax operators \cite{16}
\begin{equation}
{\mathcal{L}}_{n}=\sum_{j\in Z}u_{n-j}\partial ^{j},
\end{equation}%
allowing both positive as well as nonlocal powers of the differential $%
\partial ^{j}$. The fields $u_{j}$ of arbitrary conformal spin $j$ did not
define a primary basis. The construction of primary fields from the $u_{j}$
one's is originated from the well-known covariantization method of
Di-Francesco -Itzykson-Zuber (DIZ)\cite{17} showing that the primary $w_{j}$
fields are given by adequate polynomials of $u_{j}$ and their k-th
derivatives $u_{j}^{(k)}$.\newline
\newline
More recently there has been a growth in the study of the supersymmetric
version of conformal and $w$-symmetries in connection to integrable systems
from the point of view of field theory \cite{6, 7, 8} and through the Lax
formalism and the theory of pseudo differential operators \cite{9, 18, 19}.
Much attention has been paid also to derive the supersymmetric extension of
the Gelfand-Dickey Poisson bracket. The importance of this bracket is that
it can reproduce successfully the classical form of the superconformal (and
super w-) algebra.\newline
\newline
Besides its crucial role in string theories \cite{20} and the theory of
representation \cite{21}, the importance of Lie superlagbera in relation
with supersymmetrization and integrability is motivated by the following:
given a set of simple roots for some Lie algebra, one can construct an
associated integrable bosonic conformal Toda field theory \cite{22}. If the
algebra is finite-dimensional then the resulting theory is massless and
exhibits an extended conformal symmetry \cite{23, 24} whilst if the algebra
is of affine Kac-Moody type, then the resulting theory is massive. In
seeking to generalize this construction, it is important to stress that
there is no obvious way to supersymmetrize a given bosonic Toda theory
whilst maintaining integrability \cite{25, 26, 27}.\newline
\newline
One can, however, write down integrable Toda theories based on Lie
superalgebras which contain both bosons and fermions but which are not
supersymmetric in general. For superalgebras, unlike conventional Lie
algebras, there can exist inequivalent bases of simple roots and each of
these inequivalent bases leads to a distinct Toda theory. Each root of a
superalgebra carries a $Z_{2}$-grading which makes it either of `bosonic' or
`fermionic' type and it turns out that it is precisely those simple root
systems which are purely fermionic which give rise to supersymmetric Toda
theories .\newline
\newline
Many important aspects of integrable models with extended conformal
symmetries including the fractional supersymmetry \cite{28} and the
noncommutativity of coordinate \cite{27} are of great interest to this
study. All these aspects with some applications of the GD Poisson bracket to
non trivial symmetries and geometries will be in the center of our future
works.

\section{The General Space of differential Lax operators}

This section is devoted to a brief account of the basic properties of the
space of differential Lax operators in the bosonic case. Presently we know
that any differential operator is completely specified by a conformal spin $%
s $, $s\in {Z}$, two integers $p$ and $q=p+n$, $n\geq 0$ defining the lowest
and the highest degrees respectively and finally $(1+q-p)=n+1$ analytic
fields $u_{j}(z)$ \cite{9} We recall that the space $\mathcal{A}$ of all
local and non local differential operators admits a Lie algebra's structure
with respect to the commutator build out of the Leibnitz product. Moreover
we find that $\mathcal{A}$ splits into $3\times 2=6$ subalgebras $\mathcal{A}%
_{j+}$ and $\mathcal{A}_{j-}$, $j=0,\pm 1$ related to each others by two
types of conjugations namely the spin and the degrees conjugations. The
algebra $\mathcal{A}_{++}$ and its dual $\mathcal{A}_{--}$ are of particular
interest as they are incorporated into the construction of the
Gelfand-Dickey (G.D) Poisson bracket of $2d$-integrable models. Let us first
consider the algebra $\mathcal{A}$ of all local and non local differential
operators of arbitrary conformal spins and arbitrary degrees, one may expand
$\mathcal{A}$ as
\begin{equation}
\mathcal{A=}{\mathcal{\oplus }}_{p\leq q}\mathcal{A}^{(p,q)}\mathcal{=}{%
\mathcal{\oplus }}_{p\leq q}{\mathcal{\oplus }}_{s}\mathcal{A}%
_{s}^{(p,q)},~p,q,s\in {Z,}
\end{equation}%
where we have denoted by $(p,q)$ the lowest and thee highest degrees
respectively and by $s$ the conformal spin. The vector space $\mathcal{A}%
^{(p,q)}$ of differential operators with given degrees $(p,q)$ but undefined
spin exhibits a Lie algebra's structure with respect to the Lie bracket for $%
p\leq q\leq 1$. To be explicit, consider the space $\mathcal{A}_{s}^{(p,q)}$
of differential operators
\begin{equation}
d_{s}^{(p,q)}=\sum_{i=p}^{q}u_{s-i}(z)\partial ^{i}.
\end{equation}%
It's straightforward to check that the commutator of two operators of $%
\mathcal{A}_{s}^{(p,q)}$ is an operator of conformal spin $2s$ and degrees $%
(p,2q-1)$. Since the Lie bracket $[.,.]$ acts as
\begin{equation}
\lbrack .,.]:\mathcal{A}_{s}^{(p,q)}\times \mathcal{A}_{s}^{(p,q)}%
\longrightarrow \mathcal{A}_{2s}^{(p,2q-1)},
\end{equation}%
imposing the closure, one gets strong constraints on the spin $s$ and the
degrees parameters $(p,q)$ namely
\begin{equation}
s=0\quad \text{and}\quad p\leq q\leq 1.
\end{equation}%
From these equations we learn in particular that the spaces $\mathcal{A}%
_{0}^{(p,q)},$ $p\leq q\leq 1$ admit a Lie algebra's structure with respect
to the bracket Eq(2.3) provided that the Jacobi identity is fulfilled. This
can be ensured by showing that the Leibnitz product is associative. Indeed
given three arbitrary differential operators $%
d_{m_{1}}^{(p_{1},q_{1})},d_{m_{2}}^{(p_{2},q_{2})}$ and $%
d_{m_{3}}^{(p_{3},q_{3})}$ we find that associativity follows by help of the
identity
\begin{equation}
\sum_{l=0}^{i}\left(
\begin{array}{c}
i \\
l%
\end{array}%
\right) \left(
\begin{array}{c}
j \\
k-l%
\end{array}%
\right) =\left(
\begin{array}{c}
i+j \\
k%
\end{array}%
\right)
\end{equation}%
where $\left(
\begin{array}{c}
i \\
j%
\end{array}%
\right) $ is the usual binomial coefficient. The spaces $\mathcal{A}%
_{0}^{(p,q)},p\leq q\leq 1$ as well as the vector space $\mathcal{A}%
_{0}^{(0,1)}$ are in fact subalgebra of the Lie algebra $\mathcal{A}%
_{0}^{(-\infty ,1)}$ which can be decomposed as
\begin{equation}
\mathcal{A}_{0}^{(-\infty ,1)}=\mathcal{A}_{0}^{(-\infty ,-1)}\oplus
\mathcal{A}_{0}^{(0,1)}
\end{equation}%
$\mathcal{A}_{0}^{(-\infty ,-1)}$ is nothing but the Lie algebra of Lorentz
scalar pure pseudo-differential operators of higher degree $q=-1$ and $%
\mathcal{A}_{0}^{(0,1)}$ is the central extension of the Lie algebra $%
\mathcal{A}_{0}^{(1,1)}$ of vector fields $Diff(S^{1})$,
\begin{equation}
\mathcal{A}_{0}^{(0,1)}=\mathcal{A}_{0}^{(0,0)}\oplus \mathcal{A}_{0}^{(1,1)}
\end{equation}%
and where $\mathcal{A}_{0}^{(0,0)}$ is the one dimensional trivial ideal.
The infinite dimensional huge space $\mathcal{A}$ is the algebra of
differential operators of arbitrary spins and arbitrary degrees. It's
obtained from the space $\mathcal{A}^{(p,q)}$ by summing over all allowed
degrees
\begin{equation}
\mathcal{A=}{\mathcal{\oplus }}_{{p\leq q}}\mathcal{A}^{(p,q)}
\end{equation}%
or equivalently
\begin{equation}
\begin{array}{ccc}
\mathcal{A} & = & {\mathcal{\oplus }}_{{p\in {Z}}}[{\mathcal{\oplus }}_{\
n\in {N}}\mathcal{A}^{(p,p+n)}] \\
& = & {\mathcal{\oplus }}_{{p\in {Z}}}[{\mathcal{\oplus }}_{{n\in {N}}}[{%
\mathcal{\oplus }}_{{s\in {Z}}}\mathcal{A}_{s}^{(p,p+n)}]]%
\end{array}%
\end{equation}%
This infinite dimensional space which is the combined conformal spin and
degrees tensor algebra is closed under the Lie bracket without any
constraint. A remarkable property of $\mathcal{A}$ is that it can splits
into six infinite subalgebras $\mathcal{A}_{j+}$ and $\mathcal{A}_{j-}$, $%
j=0,\pm 1$ related to each others by conjugation of the spin and degrees.
Indeed given two integers $p$ and $q\geq p$ it is not difficult to see that
the vector spaces $\mathcal{A}^{(p,q)}$ and $\mathcal{A}^{(-q-1,-p-1)}$ are
dual with respect to the pairing product $(.,.)$ defined as
\begin{equation}
(d^{(r,s)},d^{(p,q)})=\delta _{0,1+r+q}\delta _{0,1+s+p}res[d^{(r,s)}\times
d^{(p,q)}],
\end{equation}%
where $d^{(r,s)}$ are differential operators with fixed degrees $(r,s;s\geq
r)$ but arbitrary spin and where the residue operation $res$ is defined as
\begin{equation}
res(\partial ^{i})=\delta _{0,i+1}
\end{equation}%
This equation shows that the operation $res$ exhibits a conformal spin $%
\Delta =1$. using the properties of this operation and the pairing product
eq(2.10) one can decompose $\mathcal{A}$ as follows
\begin{equation}
\mathcal{A=A}_{+}\oplus \mathcal{A}_{-}
\end{equation}%
with
\begin{equation}
\mathcal{A}_{+}={\mathcal{\oplus }}_{{p\geq 0}}[{\mathcal{\oplus }}_{{n\in {N%
}}}\mathcal{A}^{(p,p+n)}]
\end{equation}%
\begin{equation}
\mathcal{A}_{-}={\mathcal{\oplus }}_{{p\geq 0}}[{\mathcal{\oplus }}_{{n\in {N%
}}}\mathcal{A}^{(-p-n-1,-p-1)}]
\end{equation}%
The indices $+$ and $-$ carried by $\mathcal{A}_{+}$ and $\mathcal{A}_{-}$
refer to the positive (local) and negative (non local) degrees respectively.
On the other hand one can split the space $\mathcal{A}^{(p,p+n)},n\geq 0 $
as
\begin{equation}
\mathcal{A}^{(p,p+n)}=\Sigma _{-}^{(p,p+n)}\oplus \Sigma
_{0}^{(p,p+n)}\oplus \Sigma _{+}^{(p,p+n)}
\end{equation}%
$\Sigma _{-}^{(p,p+n)}$ and $\Sigma _{+}^{(p,p+n)}$ denotes the spaces of
differential operators of negative and positive definite spin. They read as
\begin{equation}
\Sigma _{-}^{(p,p+n)}={\oplus }_{{s>0}}\mathcal{A}_{-s}^{(p,p+n)}
\end{equation}%
\begin{equation}
\Sigma _{0}^{(p,p+n)}=\mathcal{A}_{0}^{(p,p+n)}
\end{equation}%
\begin{equation}
\Sigma _{+}^{(p,p+n)}={\oplus }_{{s>0}}\mathcal{A}_{s}^{(p,p+n)}
\end{equation}%
$\Sigma _{0}^{(p,p+n)}$ is just the vector space of Lorenz scalar
differential operators. Combining eqs(2.12-18) one sees that $\mathcal{A}$
decomposes into $6=3\times 2$ subalgebras
\begin{equation}
\mathcal{A}={\oplus }_{{j=0,\pm }}[\mathcal{A}_{j+}\oplus \mathcal{A}_{j-}]
\end{equation}%
with
\begin{equation}
\mathcal{A}_{j+}={\oplus }_{{p\geq 0}}[{\oplus }_{{n\in {N}}}\Sigma
_{j}^{(p,p+n)}]
\end{equation}%
\begin{equation}
\mathcal{A}_{j-}={\oplus }_{{p\geq 0}}[{\oplus }_{{n\in {N}}}\Sigma
_{j}^{(-p-n-1,-p-1)}]
\end{equation}%
The duality of these $6=3\times 2$ subalgebras is described by the combined
scalar product $\ll .,.\gg $ built out of the product eq(2.10) and conformal
spin pairing
\begin{equation}
<u_{k},u_{l}>:=\int dz~u_{k}(z)u_{1-k}(z)\delta _{k+l,1}
\end{equation}%
as follows \cite{7}.\newline
\begin{equation}
\ll d_{m}^{(r,s)},d_{n}^{(p,q)}\gg :=\delta _{0,n+m}\delta _{0,1+q+r}\delta
_{0,1+p+s}\int dz~res[d_{m}^{(r,s)}\times d_{-m}^{(-s-1,-r-1)}]
\end{equation}%
with respect to this new product; $\mathcal{A}_{++}$, $\mathcal{A}_{0+}$ and
$\mathcal{A}_{-+}$\ behaves as the dual algebras of $\mathcal{A}_{--}$, $%
\mathcal{A}_{0-}$ and $\mathcal{A}_{+-}$ respectively while $\mathcal{A}%
_{0-} $ is just the algebra of Lorenz scalar pure pseudo-operators. This
algebra and its dual $\mathcal{A}_{0+}$, the space of Lorenz scalar local
differential operators, are very special subalgebras as they are
systematically used to construct new realizations of the $w_{i}$-symmetry, $%
i\geq 2$ by using scalar differential operators type.\newline
\begin{equation}
l^{(k)}(a)=a_{-k}(a)\partial ^{k}
\end{equation}%
To close this short recapitulating section, we note that the space $\mathcal{%
A}_{++}$ is the algebra of local differential operators of positive definite
spins and positive degrees. $\mathcal{A}_{--}$ however, is the Lie algebra
of pure pseudo-differential operators of negative degrees and spins. It is
these two kinds of algebras which are usually considered in the construction
of the G.D Poisson bracket in the bosonic case as it's explicitly shown in
\cite{9}.

\section{Supersymmetric Lax operators}

\subsection{Basics definitions}

The aim of this section is to describe the supersymmetric extension of the
space of bosonic Lax operators introduced previously. This supersymmetric
generalization which is straightforward and natural in the fist steps,
exhibits some non trivial properties and make the fermionic study more
fruitful. Using the space of supersymmetric Lax operators, one can derive
the Hamiltonian structure of non linear two dimensional super integrable
models obtained by extending the bosonic Hamiltonian structure defined on
the algebra $\mathcal{A}_{++}\oplus \mathcal{A}_{--}$.\newline
Let's first consider the ring of all analytic super fields $u_{\frac{k}{2}}(%
\hat{z}),~k\in {Z}$, which depend on $(1|1)$ superspace coordinates $\hat{z}%
=(z,\theta )$. In this super commutative $Z_{2}-$graded ring $R\left[ u(\hat{%
z})\right] $, one can define an odd super derivation $D=\partial _{\theta
}+\theta \partial $, the N=1 supercovariant derivatives which obeys $N=1$
supersymmetric algebra $D^{2}=\partial $ with $\theta ^{2}=0$ and $\partial
_{\theta }\equiv \int d\theta $. Note that the supersymmetric G.D bracket,
which we shall discuss in the sequel, defines a Poisson bracket on the space
of functional of the superfields $u_{\frac{k}{2}}(\hat{z})$ defined on the
ring $R\left[ u(\hat{z})\right] $.\newline
We define the ring $\Sigma \left[ D\right] $ of differential supersymmetric
operators as polynomials in $D$ with coefficients in $R$. Using our previous
notation, one set
\begin{equation}
\Sigma \lbrack D]={\oplus }_{{m\in Z}}~{\oplus }_{{p\leq q}}\Sigma _{\frac{m%
}{2}}^{(p,q)}[D]\quad p,q\in {Z}
\end{equation}%
where $\Sigma _{\frac{m}{2}}^{(p,q)}[D]$ is the space of supersymmetric
operators type
\begin{equation}
\pounds _{\frac{m}{2}}^{(p,q)}[u]=\sum_{i=p}^{q}u_{\frac{m-i}{2}}(\hat{z}%
)D^{i}\quad p,q\in {Z}
\end{equation}%
$\Sigma _{\frac{m}{2}}^{(p,q)}$ behaves as a $(1+q-p)$ dimensional
superspace. Note also that the ring $R$ of all graded superfields can be
decomposed as
\begin{equation}
R\equiv R^{(0,0)}:={\oplus }_{{k\in Z}}R_{\frac{k}{2}}^{(0,0)}
\end{equation}%
where $R_{\frac{k}{2}}^{(0,0)}$ is the set of superfield $u_{\frac{k}{2}}(%
\hat{z})$ indexed by half integer conformal spin $\frac{k}{2},k\in Z$. Thus,
the one dimensional objects $u_{\frac{m-i}{2}}(\hat{z})D^{i}$ are typical
elements of the superspace
\begin{equation}
\Sigma _{\frac{m}{2}}^{(i,i)}=R_{\frac{m-i}{2}}^{(0,0)}\times D^{i}\equiv R_{%
\frac{m-i}{2}}^{(0,0)}\otimes \Sigma _{\frac{i}{2}}^{(i,i)}
\end{equation}%
which is fundamental in the construction of supersymmetric operators type
eq(3.2). The expression eq(3.2) means also that
\begin{equation}
\Sigma _{\frac{m}{2}}^{(p,q)}[D]\equiv {\oplus }_{{i=p}}^{q}\Sigma _{\frac{m%
}{2}}^{(i,i)}
\end{equation}%
Indeed these definitions are important in the sense that one can easily
identify all objects of the huge superspace $\Sigma $. An element $\pounds $
of $\Sigma \lbrack D]$ is called a local supersymmetric Lax operator if it
is homogeneous under the ${Z}_{2}$-grading
\begin{equation}
|x|=:%
\begin{cases}
0 & \text{for $x$ even} \\
1 & \text{otherwise}%
\end{cases}%
\end{equation}%
and have the following form at order $n,n\in {N}$%
\begin{equation}
\pounds _{\frac{n}{2}}^{(0,n)}:=\sum_{i=0}^{n}u_{\frac{i}{2}}(\hat{z})D^{n-i}
\end{equation}%
The homogeneity condition simply states that the ${Z}_{2}$-grading of the
N=1 superfield $u(\hat{z})$ is defined as
\begin{equation}
\left\vert u_{\frac{i}{2}}(\hat{z})\right\vert =i\,({mod}\,2)
\end{equation}%
The space of supersymmetric Lax operators is refereed hereafter to as $%
\Sigma _{\frac{n}{2}}^{(0,n)}$ and exhibits a dimension $n+1$. We recall
that the upstairs integers $(0,n)$ are the lowest and the highest degrees of
$\pounds $ \ and the down stair index $\frac{n}{2}$ is the spin of $\pounds $%
. To define a Lie algebraic structure on the superspace $\Sigma $ one need
to introduce a graded commutator defined for two arbitrary operators $X$ and
$Y$ as
\begin{equation}
\left[ X,Y\right] _{i}=XY-(-)^{i}YX
\end{equation}%
where the index $i=\overline{0}$ or $\overline{1}$ refer to the commutator
[,] or anticommutator \{,\} respectively. As shown in section 2, this index
is related to the graduation of the super operators $X$ and $Y$ as follows
\begin{equation}
i=\left\vert X\right\vert .\left\vert Y\right\vert
\end{equation}%
Moreover, the super commutator eq(3.9) must satisfy the
\begin{equation}
\left[ X,Y\right] _{i}=-(-)^{i}\left[ Y,X\right] _{i}
\end{equation}%
with%
\begin{equation}
\begin{array}[t]{lllll}
\left[ X,Y\right] _{\overline{0}} & = & XY-YX & = & \left[ X,Y\right] \\
\left[ X,Y\right] _{\overline{1}} & = & XY+YX & = & \left\{ X,Y\right\}%
\end{array}%
\end{equation}%
and the super-Jacobi identity
\begin{equation}
\lbrack \lbrack X,Y]_{i},Z]_{j}+[[Z,X]_{i},Y]_{j}+[[Y,Z]_{i},X]_{j}=0
\end{equation}%
Next we introduce the multiplication of operators in the superspace $\Sigma $
and derive some crucial algebraic properties. Let $\Sigma _{\frac{j}{2}%
}^{(0,0)}\equiv R_{\frac{j}{2}}^{(0,0)}$ be the ring of analytic superfields
$\phi _{\frac{j}{2}}(\hat{z})\equiv \phi (\hat{z})$ of conformal spin $\frac{%
j}{2}$ and $\Sigma _{\frac{j}{2}}^{(p,q)}$ be a superspace endowed with a
super derivative $D$ such that
\begin{equation}
D^{(l)}\left( \Sigma _{\frac{j}{2}}^{(p,q)}\right) \subseteq \Sigma _{\frac{%
j+l}{2}}^{(p,q+l)}
\end{equation}%
We will denote the superfield derivatives $\left( D\phi \right) $, $\left(
D^{2}\phi \right) $, ...,$\left( D^{i}\phi \right) $ simply as $\phi
^{\prime }$, $\phi ^{\prime \prime }$, ...,$\phi ^{(i)}$ respectively. The
multiplication of operators in $\Sigma $ is defined with respect to the
super Leibnitz rule given by the following mapping
\begin{equation}
D^{(l)}:R_{\frac{j}{2}}^{(0,0)}\longrightarrow \Sigma _{\frac{j+l}{2}%
}^{(p,l)}
\end{equation}%
such that \cite{4}%
\begin{equation}
D^{(l)}\phi \left( \hat{z}\right) =\sum_{i=0}^{\infty }\left[
\begin{array}{c}
l \\
l-i%
\end{array}%
\right] (-)^{j(l-i)}\phi ^{(i)}D^{(l-i)}
\end{equation}%
where $l$ is an arbitrary integer and the super binomial coefficients $\left[
\begin{array}{c}
l \\
k%
\end{array}%
\right] $ are defined by
\begin{equation}
\left[
\begin{array}{c}
l \\
k%
\end{array}%
\right] =:%
\begin{cases}
0 & \text{for }k>l\text{ for }(k,l)=(0,1){mod}2 \\
\left(
\begin{array}{c}
\left[ \frac{l}{2}\right] \\
\left[ \frac{k}{2}\right]%
\end{array}%
\right) & \text{otherwise}%
\end{cases}%
\end{equation}

The lowest degree $p$ of the superspace $\Sigma _{\frac{j+l}{2}}^{(p,l)}$
eq(3.15) is given by

\begin{equation}
p=:%
\begin{cases}
0 & \text{if }l\geq 0 \\
-\infty  & if\text{ \ }l\leq -1%
\end{cases}%
\end{equation}%
The symbol $[x]$ stands for the integer part of $x\in \frac{{Z}}{2}$ and $%
\left(
\begin{array}{c}
i \\
j%
\end{array}%
\right) $ is the usual binomial coefficient. The binomial and super binomial
coefficients satisfy among others, the following useful properties:\newline
a/

\begin{eqnarray}
\left(
\begin{array}{c}
p \\
q%
\end{array}%
\right)  &=&:%
\begin{cases}
1 & \text{for }q=0\text{ or }q=p \\
\frac{p(p-1)..(q+1)}{(p-q)!} & \text{for }q\prec p \\
0 & \text{otherwise}%
\end{cases}
\notag \\
q\left(
\begin{array}{c}
p \\
q%
\end{array}%
\right) (-)^{q} &=&:p\left(
\begin{array}{c}
-q \\
-p%
\end{array}%
\right) (-)^{p} \\
\left(
\begin{array}{c}
-p \\
q%
\end{array}%
\right)  &=&:(-)^{q}\left(
\begin{array}{c}
p+q-1 \\
q%
\end{array}%
\right)   \notag
\end{eqnarray}%
b/

\bigskip \qquad
\begin{eqnarray}
\left[
\begin{array}{c}
2p \\
2q\pm 1%
\end{array}%
\right]  &=&0,\qquad p,q\in {Z}  \notag \\
\left[
\begin{array}{c}
2p+1 \\
2p%
\end{array}%
\right]  &=&\left[
\begin{array}{c}
p \\
0%
\end{array}%
\right] =\left[
\begin{array}{c}
p \\
p%
\end{array}%
\right] =\left[
\begin{array}{c}
2p+1 \\
1%
\end{array}%
\right] =1 \\
\left[
\begin{array}{c}
2p \\
2q%
\end{array}%
\right]  &=&\left[
\begin{array}{c}
2p+1 \\
2q%
\end{array}%
\right] =\left[
\begin{array}{c}
2p+1 \\
2q+1%
\end{array}%
\right] =\left(
\begin{array}{c}
p \\
q%
\end{array}%
\right)   \notag \\
\left[
\begin{array}{c}
p \\
q%
\end{array}%
\right]  &=&\left( -\right) ^{\left[ \frac{q}{2}\right] }\left[
\begin{array}{c}
q-p-1 \\
q%
\end{array}%
\right]   \notag
\end{eqnarray}

\bigskip

The Leibnitz rule of local super derivatives of the superfield $\phi =\phi
_{j/2}(\hat{z})$ reads as

\bigskip
\begin{eqnarray}
D^{2i}\phi  &=&\phi ^{\left( 2i\right) }+i\phi ^{\left( 2i-2\right) }D^{2}+%
\frac{i(i-1)}{2}\phi ^{\left( 2i-4\right) }D^{4}+...  \notag \\
&=&\phi ^{\left( 2i\right) }+\sum_{n=1}^{i}\frac{i(i-1)...(i+1-n)}{n!}\phi
^{\left( 2i-2n\right) }D^{2n}
\end{eqnarray}

\bigskip
\begin{eqnarray}
D^{2i+1}\phi  &=&\sum_{k=1}^{2i+1}\left[
\begin{array}{c}
2i+1 \\
\\
k%
\end{array}%
\right] \left( -\right) ^{j(2i+1-k)}\phi ^{(k)}D^{2i+1-k}  \notag \\
&=&\sum_{p=0}^{i}\left(
\begin{array}{c}
i \\
p%
\end{array}%
\right) \left( \phi ^{(2p+1)}+\left( -\right) ^{j}\phi ^{(2p)}D\right)
D^{2i-2p},
\end{eqnarray}
where $i$ is a positive integer and $j$ is the graduation of superfield $%
\phi (\hat{z})$. Similar formulas can be written for non local super
derivatives of the superfield $\phi (\hat{z})$, we have
\begin{equation}
\begin{array}{ccc}
D^{-k}\phi  & = & \sum_{l=0}^{\infty }m_{kl}(\phi )D^{-k-l} \\
&  &  \\
m_{kl}(\phi ) & = & \left[
\begin{array}{c}
-k \\
-k-l%
\end{array}%
\right] \left( -\right) ^{j(k+l)}\phi ^{(l)}%
\end{array}%
\end{equation}%
One have then to define an infinite matrix $M(\phi )$ whose entries $%
m_{kl}(\phi )$ are functions which depend on the superfield $\phi $ and its
derivatives. The integers $k$ and $l$ indicate respectively the index of
\textquotedblright row\textquotedblright\ and \textquotedblright
coulumn\textquotedblright\ of the huge matrix $\left[ M(\phi )\right] _{kl}$
, $k=-\infty ,...,-2,-1$ and $l=0,1,2,...,\infty .$\newline
In this context, we express the non locality property of the super
derivatives $D^{-k}\phi \left( \hat{z}\right) $ by an infinite order higher
triangular matrix $\left[ M(\phi )\right] _{kl}$ which acts as follows
\begin{equation}
\left[ M\right] _{k=i,l}\equiv \left( m_{i0},m_{i1},..,..\right) :\left(
\begin{array}{c}
D^{-i} \\
D^{-i-1} \\
\vdots
\end{array}%
\right) \mapsto D^{-i}\phi
\end{equation}%
for a fixed row's index $k=i,i\geq 1$. An important aim of this formulation
is to define Poisson brackets on the superspace
\begin{equation}
{\oplus }_{{n\geq 0}}\left[ \Sigma _{\frac{n}{2}}^{(0,n)}\oplus \left[
\Sigma _{\frac{n}{2}}^{(0,n)}\right] ^{\ast }\right]
\end{equation}%
where $\left[ \Sigma _{\frac{n}{2}}^{(0,n)}\right] ^{\ast }$ is a subspace
of the super Volterra algebra of pseudo-differential operators which is dual
to $\Sigma _{\frac{n}{2}}^{(0,n)}$. We note that the algebra eq(3.25) is
just the supersymmetric analogue of the bosonic algebra $\mathcal{A}_{++}%
\mathcal{\oplus A}_{--}$ introduced previously in the construction of the
bosonic Gelfand-Dickey bracket. The functional involved in the definition of
Gelfand-Dickey(G.D) super bracket are of the form
\begin{equation}
F\left[ u\left( \hat{z}\right) \right] =\int_{B}f(u),
\end{equation}%
where $f(u)$ is an homogenous differential polynomial of the $u$'s and $%
\int_{B}$ is the well known Berezin integral $\int d\hat{z}=\int dz.d\theta $
which is usually defined in the following way: for any $u\left( \hat{z}%
\right) =a+\theta b$ and $f(u)=A(a,b)+\theta B(a,b)$ we have $%
\int_{B}f(u)=\int dzB(a,b)$. Next we introduce the notions of super-residue
and super trace which are necessary for the present study. Given a
super-pseudo operator $\mathcal{P}$ in a super Volterra basis
\begin{equation}
\mathcal{P=}\sum_{i\in {Z}}D^{i}f_{i}(\hat{z})
\end{equation}%
The super-residue is defined as
\begin{equation}
Sres\mathcal{P=}\int_{B}\left( -\right) ^{\left\vert f_{i}\right\vert
}f_{-1}(\hat{z})
\end{equation}%
Note that the residue operation (res) introduced in the bosonic case
eq(2.11), exhibits a spin $\Delta (res)=1$, while the spin of the
super-residue operation $\Delta (Sres)=\frac{1}{2}$, fact which is immediate
if we remark that $\Delta (d\hat{z})=\Delta (dzd\theta )=\frac{-1}{2}$. One
can also easily show that the super residue of a graded supercommutator is a
total derivative, so that its supertrace is a vanishing number
\begin{equation}
Sres\left[ \pounds ,P\right\} =total\,\,\,derivative
\end{equation}%
\begin{equation}
Str\left[ \pounds ,P\right\} =0,
\end{equation}%
for every $\pounds \in \Sigma $ and $P\in \Sigma ^{\ast }$. Since $\left[
\pounds ,R\right\} =\pounds P-\left( -\right) ^{\left\vert \pounds %
\right\vert \left\vert P\right\vert }P\pounds $, the property eq(3.30) means
that it's possible to define a graded superbilinear form
\begin{equation}
Str\left( \pounds P\right) =\left( -\right) ^{\left\vert \pounds \right\vert
\left\vert P\right\vert }Str\left( P\pounds \right)
\end{equation}%
on the superspace
\begin{equation}
\Sigma _{++}\oplus \Sigma _{--}={\oplus }_{{n\geq 0}}\left[ \Sigma _{\frac{n%
}{2}}^{(0,n)}\oplus \left[ \Sigma _{\frac{n}{2}}^{(0,n)}\right] ^{\ast }%
\right]
\end{equation}%
This form pairs the super differential operators $\pounds $ of $\Sigma _{++}$
and the super pseudo-operators $P$ of $\Sigma _{--}$ as follows
\begin{equation}
\ll \pounds _{\frac{n}{2}}^{(0,n)},P_{\frac{m}{2}}^{(r,s)}\gg :=\delta
_{n+m,0}\delta _{n+r+1,0}\delta _{s+1,0}\int_{B}Sres\left( \pounds _{\frac{n%
}{2}}^{(0,n)}\circ P_{\frac{-n}{2}}^{(-n-1,-1)}\right)
\end{equation}%
This supersymmetric scalar product, which connect both the super-residue and
degrees pairing, can be rewriting in a similar form
\begin{equation}
\ll \pounds _{\frac{n}{2}}^{(0,n)},P_{\frac{-n}{2}}^{(-n-1,-1)}\gg
=\int_{B}\sum_{i=1}^{n}\left( -\right) ^{i+1}\left( u_{\frac{i}{2}}\left(
\hat{z}\right) \chi _{\frac{1-i}{2}}\left( \hat{z}\right) \right) ,
\end{equation}%
where
\begin{equation}
\pounds _{\frac{n}{2}}^{(0,n)}=\sum_{i=0}^{n}u_{\frac{i}{2}}\left( \hat{z}%
\right) D^{n-i}
\end{equation}%
\begin{equation}
P_{\frac{-n}{2}}^{(-n-1,-1)}=\sum_{i=1}^{n+1}D^{-i}\chi _{\frac{i-n}{2}%
}\left( \hat{z}\right)
\end{equation}%
The supersymmetric Lax operators usually are those for which $u_{0}\left(
\hat{z}\right) =1$. This simple choice which is consistent with the
definition of (supersymmetric) $w$-symmetries imply a constraint on the
corresponding dual superfield $\chi _{\frac{1}{2}}\left( \hat{z}\right) $,
namely
\begin{equation}
\chi _{\frac{1}{2}}\left( \hat{z}\right) =0
\end{equation}%
On the other hand, using the supersymmetric combined product eq(3.34), it is
not difficult to see that
\begin{equation}
\left[ \Sigma _{\frac{n}{2}}^{(0,n)}\right] ^{\ast }=\Sigma _{\frac{-n}{2}%
}^{(-n-1,-1)}
\end{equation}%
So, the super Lax operator $\pounds $ and its dual $P$ eqs(3.35-36) read
\begin{equation}
\pounds =D^{n}+\sum_{i=1}^{n}u_{\frac{i}{2}}\left( \hat{z}\right) D^{n-i}
\end{equation}%
\begin{equation}
P=\sum_{i=1}^{n}D_{{}}^{-i}\chi _{\frac{i-n}{2}}\left( \hat{z}\right)
\end{equation}%
For the formal sum eq(3.36), note that only a finite number of the
superfields $\chi _{\frac{i-n}{2}}\left( \hat{z}\right) $ are non zero. The
super-residue duality imply that it is possible to realize the
pseudo-superfields $\chi _{\frac{i}{2}}\left( \hat{z}\right) $ in terms of $%
u_{\frac{j}{2}}\left( \hat{z}\right) $. Indeed, let us consider a functional
$f[u_{\frac{1}{2}},u_{1},u_{\frac{3}{2}},...,u_{n}]$ acting on the ring $R_{%
\frac{k}{2}}^{(0,0)}$ of chiral super fields $u_{\frac{k}{2}}\left( \hat{z}%
\right) $. We have
\begin{equation}
P_{f}=\sum_{j=1}^{n}D^{-j}.\chi _{\frac{j-n}{2}}\left( \hat{z}\right) ,
\end{equation}%
with
\begin{equation}
\chi _{\frac{1-k}{2}}\left( \hat{z}\right) :=\left( -\right) ^{k}\frac{%
\delta f\left[ u\right] }{\delta u_{\frac{k}{2}}},k=1,2,...
\end{equation}%
\begin{equation}
\Delta \left( \frac{\delta f}{\delta u_{\frac{k}{2}}}\right) =\frac{1-k}{2}
\end{equation}%
Note by the way that, in addition to the functionals $f\left[ u\right] ,$
other geometrical objects that are necessary to construct the super
symmetric Gelfand-Dickey brackets are given by vector fields (1-forms) and
the map sending a function to its associated Hamiltonian vector field, ie
the coadjoint supersymmetric operator . Here we will not need to follow this
procedure, however we shall concentrate in the next part of this work on the
infinitesimal (coadjoint) supersymmetric operator, which is fundamental in
the definition of G.D super bracket and the derivation of higher spin
extensions of the conformal symmetry.

\section{The super-Hamiltonian operator $V_{P}(L):$}

First of all we remind that the supersymmetric G.D Poisson bracket is of the
form
\begin{equation}
\left\{ F[u_{\frac{i}{2}}]~,G[u_{j}]\right\} =\int_{B}Sres\left\{
V_{P_{F}}(L)\circ P_{G}\right\} ,
\end{equation}%
where $F$ and $G$ are functionals of the superfield $u_{i}(\hat{z}).$ This
definition of the super G.D bracket is based on the super-residue duality
eq(3.30) of super differential operators $L$ and super volterra ones $P$.
The map which combines these two kind of dual operators is given by the
hamiltonian operator:
\begin{equation}
V_{P}(L)=L(PL)_{+}-(LP)_{+}L\text{ ,\ }
\end{equation}%
where the subscripts $+$ indicate the restriction to the local part, ie
\begin{equation}
\left[ \sum\limits_{i\epsilon Z}a_{i}(\hat{z})D^{i}\right] _{+}=\sum_{i\geq
0}a_{i}(\hat{z})D^{i}
\end{equation}%
Before one turns to the supersymmetric analysis of the G.D algebra and its
induced superconformal and super $w$-symmetry, we give here below an
explicit description of the hamiltonian operator $V_{P}(L)$ which is defined
as the infinitesimal action of super- pseudo operators $P$ on the space of
supersymmetric Lax operators $L$. In order to simplify the notation we have
considered $L_{n/2}^{(0,n)}\equiv L$ and $P_{-n/2}^{(-n-1,-1)}\equiv P$ . An
important property of $V_{P}(L)$ is its fundamental role in describing both
the first and the second hamiltonian G.D Poisson brackets occurring in the
definition of super integrable systems. Knowing that the first G.D bracket
is constructed by using the local graded commutator $\left[ P,L\right] _{+}$
while the second G.D bracket eq(4.1) is generated by $V_{P}(L)$ , the
following shift for example:
\begin{equation}
L\longrightarrow L+\lambda \equiv \tilde{L},
\end{equation}%
shows clearly the relevance of the second GD bracket. Indeed remark that $%
V_{P}(L)$ transform with respect to eq(4.4) like:
\begin{equation}
V_{P}(L)\longrightarrow V_{P}(L)+\lambda \left[ P,L\right] _{+}=V_{P}(\tilde{%
L}),
\end{equation}%
showing in turn how the first GD bracket can be described by the second one.
Next we focus to work out supersymmetric hamiltonian operators $V_{P}(L)$
for super Lax operators $L$ of degree three and five.

\subsection{The $N=2$ superconformal algebra.}

Let's consider the following super (pseudo) operators
\begin{equation}
L=D^{3}+UD^{2}+VD+W,\text{ \ \ \ \ \ \ \ \ \ \ \ \ \ \ \ \ }L\in \Sigma _{%
\frac{3}{2}}^{(0,3)}
\end{equation}%
\begin{equation}
P=D^{-1}X+D^{-2}Y+D^{-3}Z,\text{ \ \ \ \ \ \ \ \ \ \ \ \ \ \ \ \ \ \ \ }P\in
\Sigma _{-\frac{3}{2}}^{(-3,-1)}
\end{equation}%
where $\left( U,V,W\right) $ and $\left( X,Y,Z\right) $ are superfields of
spin $\left( \frac{1}{2},1,\frac{3}{2}\right) $ and $\left( -1,-\frac{1}{2}%
,0\right) $ respectively. The superfields are constrained by the super
residue duality
\begin{eqnarray}
Sres\left( L.P\right)  &=&\sum_{i=1}^{3}(-)^{i+1}u_{\frac{i}{2}}(\hat{z})v_{%
\frac{1-i}{2}}(\hat{z})  \notag \\
&&  \notag \\
&=&UZ-VY+WX
\end{eqnarray}%
with the convention notation $\left( u_{\frac{1}{2}},u_{1,}u_{\frac{3}{2}%
}\right) \equiv \left( U,V,W\right) $ and $\left( v_{-1},v_{-\frac{1}{2}%
},v_{0}\right) \equiv \left( X,Y,Z\right) .$ Note that we have to set $%
U\equiv u_{\frac{1}{2}}(\hat{z})=0$ which is the traceless condition
required by the $\ sl(2|2)$ Lie super algebra structure. This condition is
equivalent to the following coset superspace operation
\begin{equation}
\Sigma _{\frac{3}{2}}^{(0,3)}\text{ \ }\longrightarrow \text{ \ \ }\Sigma _{%
\frac{3}{2}}^{(0,3)}\text{ }/\text{ }\Sigma _{\frac{3}{2}}^{(2,2)}
\end{equation}%
where $\Sigma _{\frac{3}{2}}^{(2,2)}$ is the one dimensional subspace of $%
\Sigma _{\frac{3}{2}}^{(0,3)}$ which is generated by half spin superfield $%
U\equiv u_{\frac{1}{2}}(\hat{z}).$ At first sight, it seems difficult how to
determine the superfield $Z\equiv $ $v_{0}$ of spin zero, dual to the
vanishing superfield $U\equiv u_{\frac{1}{2}}$. To do this, one must compute
$V_{P}(L)$ for
\begin{eqnarray}
L &=&D^{3}+VD+W  \notag \\
P &=&D^{-1}X+D^{-2}Y+D^{-3}Z
\end{eqnarray}%
and require that $L$ is invariant under the coadjoint action eq(4.2).
Straightforward computations lead to
\begin{equation}
\left( LP\right) _{+}=XD^{2}-YD+X^{^{\prime \prime }}+Y^{^{\prime }}+VX+Z%
\text{ \ }
\end{equation}%
\begin{equation*}
\end{equation*}%
\begin{equation}
\left( PL\right) _{+}=XD^{2}+(X^{^{\prime }}+Y)D-X^{^{\prime \prime }}+XV+Z,
\end{equation}%
implying in turns the following
\begin{equation}
L\left( P.L\right) _{+}=\sum_{i=0}^{5}A_{i}(\hat{z})D^{i}\text{ \ \ }
\end{equation}%
with
\begin{equation}
\begin{array}{l}
A_{5}(\hat{z})=X \\
\\
A_{4}(\hat{z})=-Y \\
\\
A_{3}(\hat{z})=2VX+Z+Y^{^{\prime }}+X^{^{\prime \prime }} \\
\\
A_{2}(\hat{z})=WX+X^{^{\prime }}V+XV^{^{\prime }}-VY+Z^{^{\prime
}}-Y^{^{\prime \prime }}-X^{^{\prime \prime \prime }} \\
\\
A_{1}(\hat{z})=WX^{^{\prime }}+WY+VY^{^{\prime }}+VZ+XV^{2}+X^{^{\prime
\prime }}V \\
\text{ \ \ \ \ \ \ \ \ \ \ \ \ }+XV^{^{\prime \prime }}+Z^{^{\prime \prime
}}+Y^{^{\prime \prime \prime }} \\
\\
A_{0}(\hat{z})=WXV+VXV^{^{\prime }}+VX^{^{\prime }}V+WZ-WX^{^{\prime \prime
}} \\
\text{ \ \ \ \ \ \ \ \ \ \ \ \ }+XV^{^{\prime \prime \prime }}+X^{^{\prime
\prime }}V^{^{\prime }}+X^{^{\prime }}V^{^{\prime \prime }}+VZ^{^{\prime
}}+Z^{^{\prime \prime \prime }}-X^{(5)}%
\end{array}%
\end{equation}%
where
\begin{equation}
\Delta \left( A_{i}(\hat{z})\right) =\frac{3-i}{2},i=0,1,...,5
\end{equation}%
Similar computations give
\begin{equation}
\left( LP\right) _{+}L=\sum_{i=0}^{5}B_{i}(\hat{z})D^{i}
\end{equation}%
with
\begin{equation}
\begin{array}{lll}
B_{5}(\hat{z}) & = & X \\
B_{4}(\hat{z}) & = & -Y \\
B_{3}(\hat{z}) & = & 2VX+Z+Y^{^{\prime }}+X^{^{\prime \prime }}\text{ } \\
B_{2}(\hat{z}) & = & XW-YV \\
B_{1}(\hat{z}) & = & XV^{2}+ZV+YW-YV^{^{\prime }}+Y^{^{\prime
}}V+X^{^{\prime \prime }}V+XV^{^{\prime \prime }} \\
B_{0}(\hat{z}) & = & VXW+Y^{^{\prime }}W-W^{^{\prime }}Y+X^{^{\prime \prime
}}W+XW^{^{\prime \prime }}+ZW%
\end{array}%
\end{equation}%
we find
\begin{equation}
\begin{array}{ccc}
V_{P}(L) & = & \sum_{i=0}^{2}\left( A_{i}-B_{i}\right) (\hat{z})D^{i}\text{ }
\\
&  &  \\
& = & \sum_{i=0}^{2}C_{i}(\hat{z})D^{i}%
\end{array}%
\end{equation}%
with
\begin{equation}
\begin{array}{lll}
C_{0}(\hat{z}) & = &
\begin{array}[t]{l}
VXV^{^{\prime }}+VX^{^{\prime }}V-2WX^{^{\prime \prime }}+VZ^{^{\prime }} \\
+X^{^{\prime }}V^{^{\prime \prime }}+XV^{^{\prime \prime \prime
}}+X^{^{\prime \prime }}V^{^{\prime }}-Y^{^{\prime }}W \\
+W^{^{\prime }}Y-XW^{^{\prime \prime }}+Z^{^{\prime \prime \prime }}-X^{(5)}%
\end{array}
\\
C_{1}(\hat{z}) & = & WX^{^{\prime }}+2WY-YV^{^{\prime }}+Y^{^{\prime \prime
\prime }}+Z^{^{\prime \prime \prime }} \\
C_{2}(\hat{z}) & = & X^{^{\prime }}V+XV^{^{\prime }}+Z^{^{\prime
}}-Y^{^{\prime \prime }}-X^{^{\prime \prime \prime }}%
\end{array}%
\end{equation}%
It is important to remark that the operator $V_{P}(L)$ don't preserve the
Lie algebra's structure of the supersymmetric coset space eq (4.9) generated
by the super Lax operators $L$ eq(4.10). In other words, $V_{P}(L)$ is not
an affine $sl\left( 2|2\right) ^{(1)}$ operator because its $D^{2}$-term: $%
C_{2}(\hat{z})$ does not vanish in general. However, knowing that $V_{P}(L)$
share with $L$ the properties of locality (positive degrees), grading and
spin $\frac{3}{2}$\ , we can require that
\begin{equation}
C_{2}(\hat{z})=0
\end{equation}%
or
\begin{equation}
Z(\hat{z})=X^{^{\prime \prime }}+Y^{^{\prime }}-XV
\end{equation}%
Note also that the dual constraint eq(4.8) which describes the trace zero is
equivalent to:
\begin{equation}
Sres\left\{ L,P\right\} =0\text{ }
\end{equation}%
Injecting the constraint equation eq(4.21) into the expressions of $C_{0}(%
\hat{z})$ and $C_{1}(\hat{z})$ one finds
\begin{equation}
\begin{array}{lll}
C_{0}(\hat{z}) & = & Y^{(4)}+VY^{^{\prime \prime }}-2WX^{^{\prime \prime
}}-XW^{^{\prime \prime }} \\
&  &  \\
C_{1}(\hat{z}) & = & X^{(4)}+2Y^{^{\prime \prime \prime }}-(XV)^{^{\prime
\prime }}+WX^{^{\prime }}+2WY-YV^{^{\prime }}\text{ }%
\end{array}%
\end{equation}%
which imply in turns
\begin{equation}
\begin{array}{lll}
V_{P}(L) & = & \left[ -\left( \frac{\delta f}{\delta W}\right)
^{^{(4)}}+2\left( \frac{\delta f}{\delta V}\right) ^{^{\prime \prime \prime
}}+\left( V.\frac{\delta f}{\delta W}\right) ^{^{\prime \prime }}\right.  \\
&  & \left. -W\left( \frac{\delta f}{\delta W}\right) ^{^{\prime }}+2W\frac{%
\delta f}{\delta V}-\frac{\delta f}{\delta V}.V^{^{\prime }}\right] D\text{\
} \\
&  & +\left[ \left( \frac{\delta f}{\delta V}\right) ^{^{(4)}}+V\left( \frac{%
\delta f}{\delta V}\right) ^{^{\prime \prime }}\right.  \\
&  & \left. +2W\left( \frac{\delta f}{\delta W}\right) ^{^{\prime \prime }}+%
\frac{\delta f}{\delta W}.W^{^{\prime \prime }}\right]
\end{array}%
\end{equation}%
where: $\ \ \ \ \ \ X=\frac{\delta f}{\delta W}$\ and \ $Y=\frac{\delta f}{%
\delta V}$\ \ \ for $f=f\ [U,V,W].$\newline
The super Gelfand-Dickey algebra of the second kind takes then the following
form:
\begin{equation}
\begin{array}{lll}
\left\{ f\left( \hat{z}\right) ,g\left( \hat{z}\right) \right\}  & = &
\int_{B}Sres\left[ V_{P_{f}}\left( L\right) .P_{g}\right]  \\
&  &  \\
& = & \int d\hat{\sigma}\left\{ \left[ \left( \partial _{\sigma }^{2}\frac{%
\delta f}{\delta W}\right) -2\left( D\partial _{\sigma }.\frac{\delta f}{%
\delta V}\right) -\left( \partial _{\sigma }(V.\frac{\delta f}{\delta W}%
)\right) \right. \right. -W\left( D\frac{\delta f}{\delta W}\right)  \\
&  &  \\
& - & \left. 2W.\frac{\delta f}{\delta V}+\frac{\delta f}{\delta V}.(DV)%
\right] \frac{\delta g}{\delta V}-\left[ \left( \partial _{\sigma }^{2}\frac{%
\delta f}{\delta V}\right) +V\left( \partial _{\sigma }\frac{\delta f}{%
\delta V}\right) +2W\left( \partial _{\sigma }\frac{\delta f}{\delta W}%
\right) \right.  \\
&  &  \\
& + & \left. \left. +\frac{\delta f}{\delta W}\left( \partial _{\sigma
}W\right) \right] \frac{\delta g}{\delta W}\right\}
\end{array}%
\end{equation}%
Before we give the super G.D poisson bracket of the superfields $V\left(
z,\theta \right) $ of conformal spin $1$ and $\frac{3}{2}$ respectively,
let's first discuss their covariantization. In this example we have used the
$N=1$ supersymmetry to derive the $N=1$ superconformal algebra. The latter
is best described in an $N=1$ superspace with local analytic coordinates $%
\left( z,\theta \right) $. The superanalytic map is given by
\begin{equation}
\hat{z}=\left( z,\theta \right) \longrightarrow \widetilde{\hat{z}}=\left(
\tilde{z}(z,\theta )\text{ };\text{ }\tilde{\theta}(z,\theta )\right) ,
\end{equation}%
with \ $\tilde{\theta}^{2}=\theta ^{2}=0$. The $N=1$ superderivative
transforms with respect this map like:
\begin{equation}
D=\left( D\tilde{z}-\tilde{\theta}D\tilde{\theta}\right) \tilde{D}%
^{2}+\left( D\tilde{\theta}\right) \tilde{D}
\end{equation}%
The superanalytic map defines a superconformal transformation if the
superderivative $D$ transforms homogenously
\begin{equation}
D=\left( D\tilde{\theta}\right) \tilde{D}
\end{equation}%
which is equivalent also to the following constraint equation
\begin{equation}
D\tilde{z}=\tilde{\theta}D\tilde{\theta}
\end{equation}%
The transformation of the super Lax operators $L\equiv L_{\frac{1}{2}%
}^{(0,n)}$ with respect to the superconformal transformation is given by:
\begin{equation}
L\text{ \ }\longrightarrow \text{ \ }\tilde{L}=\left( D\tilde{\theta}\right)
^{\frac{-n-1}{2}}L\left( D\tilde{\theta}\right) ^{\frac{1-n}{2}}
\end{equation}%
This transformation is important in the sense that it allows to determine
the correct transformations of all the superfields $u_{\frac{i}{2}}(\hat{z})$%
. The present example gives
\begin{equation}
\tilde{D}^{3}+\tilde{V}\tilde{D}+\tilde{W}=\left( D\tilde{\theta}\right)
^{-2}\left[ D^{3}+VD+W\right] \left( D\tilde{\theta}\right) ^{-1}
\end{equation}%
identifying both sides in eq(4.31) we obtain the following transformations
laws for the superfields $V\left( z,\theta \right) $ and $W\left( z,\theta
\right) ,$
\begin{equation}
\begin{array}{lll}
V\left( z,\theta \right)  & = & \tilde{V}\left( \tilde{z},\tilde{\theta}%
\right) \left( D\tilde{\theta}\right) ^{2} \\
W\left( z,\theta \right)  & = & \tilde{W}\left( \tilde{z},\tilde{\theta}%
\right) \left( D\tilde{\theta}\right) ^{3}+\tilde{V}.D\tilde{\theta}.D^{2}%
\tilde{\theta}+S\left( \tilde{z},\tilde{z}\right)
\end{array}%
\end{equation}%
where $S\left( \tilde{z},\tilde{z}\right) $ is the super Schwarzian
derivative given by
\begin{equation}
S\left( \tilde{z},\tilde{z}\right) =\frac{\partial ^{2}\tilde{\theta}}{D%
\tilde{\theta}}-2\frac{D\partial \tilde{\theta}}{D\tilde{\theta}}\frac{%
\partial \tilde{\theta}}{D\tilde{\theta}}
\end{equation}%
The superfield $V\left( z,\theta \right) $ transforms covariantly as a field
of conformal spin one, while $W\left( z,\theta \right) $ does not have the
correct transformation property for a field with conformal dimension $\frac{3%
}{2}.$ If we consider the redefinition
\begin{equation}
W\left( z,\theta \right) \text{ \ }\longrightarrow \text{ \ }\widehat{W}%
\left( z,\theta \right) =W\left( z,\theta \right) -\frac{1}{2}\left(
DV\left( z,\theta \right) \right)
\end{equation}%
We can easily check that $\widehat{W}\left( z,\theta \right) $ transforms
covariantly as a field of conformal spin $\frac{3}{2}.$ Therefore we can
identify $V\left( z,\theta \right) $ and $\widehat{W}\left( z,\theta \right)
$ with
\begin{equation}
\begin{array}{ccc}
V\left( z,\theta \right)  & \equiv  & \Gamma \left( z,\theta \right)
=J(z)+\theta G_{1}(z) \\
\widehat{W}\left( z,\theta \right)  & \equiv  & \Sigma \left( z,\theta
\right) =G_{2}(z)+\theta T(z)%
\end{array}%
\end{equation}%
By virtue of the second hamiltonian super GD bracket, the fields $J(z)$ , $%
G_{1}(z)$ , $G_{2}(z)$ and $T(z)$ form then an $N=2$ supermultiplet $\left(
1,(\frac{3}{2})^{2},2\right) $ and satisfy an $N=2$ supersymmetric algebra
that we can write in the compact form as \bigskip
\begin{eqnarray}
\left\{ \Gamma (z,\theta ),\Gamma (z^{^{\prime }},\theta ^{^{\prime
}})\right\}  &=&-2\left( D^{^{\prime }}\partial _{z^{^{\prime }}}\Delta
\right) -2\Sigma (z^{^{\prime }},\theta ^{^{\prime }})\Delta   \notag \\
\left\{ \Sigma (z,\theta ),\Gamma (z^{^{\prime }},\theta ^{^{\prime
}})\right\}  &=&-\Gamma (z^{^{\prime }},\theta ^{^{\prime }})\left(
D^{^{\prime }2}\Delta \right) -\frac{1}{2}(D^{^{\prime }}\Gamma
)(D^{^{\prime }}\Delta )+(D^{^{\prime }2}\Gamma )\Delta  \\
\left\{ \Sigma (z,\theta ),\Sigma (z^{^{\prime }},\theta ^{^{\prime
}})\right\}  &=&-\frac{1}{2}(D^{^{\prime }5}\Delta )-\frac{3}{2}\Sigma
(z^{^{\prime }},\theta ^{^{\prime }})\left( D^{^{\prime }2}\Delta \right) -%
\frac{1}{2}(D^{^{\prime }}\Sigma )(D^{^{\prime }}\Delta )  \notag \\
&&-(D^{^{\prime }2}\Sigma )\Delta   \notag
\end{eqnarray}%
\bigskip with $\Delta =\delta (z-z^{^{\prime }}).(\theta -\theta ^{^{\prime
}})$ \ and $\ D^{^{\prime }}=\partial _{\theta ^{\prime }}+\theta ^{^{\prime
}}\partial _{z^{\prime }}$ \ .

\subsection{The $N=2$\ super $W_{3}$-algebra}

Here we describe the infinitesimal coadjoint operator $V_{P}(L)$ associated
to $N=2$ super W$_{3}$-algebra which is an extension of the supersymmetric $%
N=2$ Virasoro algebra. The affine graded superalgebra considered is $%
sl\left( 3\mid 3\right) ^{(1)}$ generated by the superfields $\left(
J,Q,T,W\right) \equiv \left( U_{1},U_{\frac{3}{2}},U_{2},U_{\frac{5}{2}%
}\right) $. These fields are the coefficients of the following five order's
super Lax operator
\begin{equation}
\begin{array}{c}
L=D^{5}+JD^{3}+QD^{2}+TD+W \\
\\
L\in \Sigma _{\frac{5}{2}}^{(0,5)}/\Sigma _{\frac{5}{2}}^{(4,4)}%
\end{array}%
\end{equation}%
The superpseudo operator corresponding to $L$ is
\begin{equation}
\begin{array}{c}
P=D^{-5}X_{5}+D^{-4}X_{4}+D^{-3}X_{3}+D^{-2}X_{2}+D^{-1}X_{1} \\
\\
P\in \Sigma _{-\frac{5}{2}}^{(-5,-1)}%
\end{array}%
\end{equation}%
with $\Delta (X_{i})=\frac{i-5}{2}$ \ , \ $i=1,2,...,5$ and $\left\vert
X_{i}\right\vert =(i+1)$ mod $2$ . The functional realization of the pseudo
superfields is given by
\begin{equation}
X_{1}=-\frac{\delta f}{\delta W},X_{2}=\frac{\delta f}{\delta T},X_{3}=-%
\frac{\delta f}{\delta Q},X_{4}=\frac{\delta f}{\delta J}
\end{equation}%
where $f=f$ $\left[ R,Q,T,W\right] .$ The scalar superfield $X_{5}(\hat{z})$
dual to the vanishing coefficient $U_{\frac{1}{2}}(\hat{z})$ of the $D^{4}$%
-term of $L$ is requested to satisfy the traceless condition
\begin{equation}
Sres\left\{ L,P\right\} =0
\end{equation}%
with
\begin{equation}
Sres\left( L,P\right) =-JX_{4}+QX_{3}-TX_{2}+WX_{1}
\end{equation}%
After a long computation we find
\begin{eqnarray}
\left( L\circ P\right) _{+}L &=&\sum_{i=0}\Gamma _{i}D^{(i)}L \\
&=&\sum_{i=0}^{9}\gamma _{i}D^{i}
\end{eqnarray}%
where $\Gamma _{i}$ , $i=0,...,4$ are superfields of spin $\Delta (\Gamma
_{i})=-\frac{i}{2}$ expressed in term of the pseudo-superfields $X_{i}$ as
follows

\begin{eqnarray*}
\Gamma _{0} &=&-J\left( \frac{\delta f}{\delta W}\right) ^{^{\prime \prime
}}+J\left( \frac{\delta f}{\delta T}\right) ^{^{\prime }}-J\left( \frac{%
\delta f}{\delta Q}\right) +Q\left( \frac{\delta f}{\delta T}\right)
-T\left( \frac{\delta f}{\delta W}\right) -\left( \frac{\delta f}{\delta W}%
\right) ^{(4)} \\
&&+\left( \frac{\delta f}{\delta T}\right) ^{^{^{\prime \prime \prime
}}}-\left( \frac{\delta f}{\delta Q}\right) ^{^{\prime \prime }}+\left(
\frac{\delta f}{\delta R}\right) ^{^{\prime }}+X_{5}-Q\left( \frac{\delta f}{%
\delta W}\right) ^{^{\prime }}
\end{eqnarray*}

\begin{eqnarray*}
\Gamma _{1} &=&-J\frac{\delta f}{\delta T}-Q\frac{\delta f}{\delta W}-\left(
\frac{\delta f}{\delta T}\right) ^{^{\prime \prime }}-\frac{\delta f}{\delta
R} \\
\Gamma _{2} &=&-J\left( \frac{\delta f}{\delta W}\right) -2\left( \frac{%
\delta f}{\delta W}\right) ^{^{\prime \prime }}+\left( \frac{\delta f}{%
\delta T}\right) ^{^{\prime }}-\frac{\delta f}{\delta Q} \\
\Gamma _{3} &=&-\frac{\delta f}{\delta T} \\
\Gamma _{4} &=&-\frac{\delta f}{\delta W}
\end{eqnarray*}
Explicit computations lead to
\begin{equation}
\begin{array}{lll}
\gamma _{0} & = & \Gamma _{0}W+\Gamma _{1}W^{^{\prime }}+\Gamma
_{2}W^{^{^{\prime \prime }}}+\Gamma _{3}W^{^{^{\prime \prime \prime
}}}+\Gamma _{4}W^{(4)} \\
\gamma _{1} & = & \Gamma _{0}T+\Gamma _{1}\left( T^{^{\prime }}-W\right)
+\Gamma _{2}T^{^{^{\prime \prime }}}+\Gamma _{3}\left( T^{^{^{\prime \prime
\prime }}}-W^{^{^{\prime \prime }}}\right) +\Gamma _{4}T^{(4)} \\
\gamma _{2} & = & \Gamma _{0}Q+\Gamma _{1}\left( Q^{^{\prime }}+T\right)
+\Gamma _{2}\left( Q^{^{^{\prime \prime }}}+W\right) +\Gamma _{3}\left(
Q^{^{^{\prime \prime \prime }}}+W^{^{^{\prime }}}+T^{^{^{\prime \prime
}}}\right)  \\
& + & \Gamma _{4}\left( Q^{(4)}+2W^{^{^{\prime \prime }}}\right)  \\
\gamma _{3} & = & \Gamma _{0}J+\Gamma _{1}\left( J^{^{\prime }}-Q\right)
+\Gamma _{2}\left( J^{^{^{\prime \prime }}}+T\right) +\Gamma _{3}\left(
J^{^{^{\prime \prime \prime }}}-Q^{^{^{\prime \prime }}}+T^{^{^{\prime
}}}-W\right)  \\
& + & \Gamma _{4}\left( J^{(4)}+2T^{^{^{\prime \prime }}}\right)  \\
\gamma _{4} & = & \Gamma _{1}J+\Gamma _{2}Q+\Gamma _{3}\left( J^{^{^{\prime
\prime }}}+Q^{^{^{\prime }}}+T\right) +\Gamma _{4}\left( 2Q^{^{^{\prime
\prime }}}+W\right)  \\
\gamma _{5} & = & \Gamma _{0}+\Gamma _{2}J+\Gamma _{3}\left( J^{^{^{\prime
}}}-Q\right) +\Gamma _{4}\left( T+2J^{^{^{\prime \prime }}}\right)  \\
\gamma _{6} & = & \Gamma _{1}+\Gamma _{3}J+\Gamma _{4}Q \\
\gamma _{7} & = & \Gamma _{2}+\Gamma _{4}J \\
\gamma _{8} & = & \Gamma _{3} \\
\gamma _{9} & = & \Gamma _{4}%
\end{array}%
\end{equation}%
with $\ \ \Delta (\gamma _{i})=\frac{5-i}{2}$. On the other hand the
explicit expressions of $L.\left( P.L\right) _{+}$ is determined by similar
calculations. We find:
\begin{equation}
\left( P.L\right) _{+}=\sum_{i=0}^{4}\lambda _{i}D^{i}
\end{equation}%
where $\lambda _{i}$ are superfields of spin $\Delta (\lambda _{i})=\frac{-i%
}{2}$ given by:

\begin{eqnarray}
\lambda _{0} &=&-\left( \frac{\delta f}{\delta W}\right) ^{(4)}+\frac{\delta
f}{\delta W}\left( J^{^{\prime \prime }}-Q^{^{\prime }}-T\right) -\left(
\frac{\delta f}{\delta W}\right) ^{^{\prime }}Q+\left( \frac{\delta f}{%
\delta W}\right) ^{^{\prime \prime }}J  \notag \\
&&+\frac{\delta f}{\delta T}.Q-\frac{\delta f}{\delta Q}.J+2\left( \frac{%
\delta f}{\delta Q}\right) ^{^{\prime \prime }}+X_{5}  \notag
\end{eqnarray}
\begin{eqnarray}
\lambda _{1} &=&\left( \frac{\delta f}{\delta W}\right) ^{^{\prime \prime
\prime }}-\left( \frac{\delta f}{\delta T}\right) ^{^{\prime \prime
}}-\left( \frac{\delta f}{\delta W}.J\right) ^{^{\prime }}-\left( \frac{%
\delta f}{\delta Q}\right) ^{^{\prime }}+\frac{\delta f}{\delta W}.Q+\frac{%
\delta f}{\delta T}.J+\frac{\delta f}{\delta J} \\
\lambda _{2} &=&\left( \frac{\delta f}{\delta W}\right) ^{^{\prime \prime }}-%
\frac{\delta f}{\delta W}.J-\frac{\delta f}{\delta Q}  \notag \\
\lambda _{3} &=&-\left( \frac{\delta f}{\delta W}\right) ^{^{\prime }}+\frac{%
\delta f}{\delta T}  \notag \\
\lambda _{4} &=&-\frac{\delta f}{\delta W}  \notag
\end{eqnarray}%
One can then easily check that $L.\left( P.L\right) _{+}$ is of the form
\begin{eqnarray}
L.\left( P.L\right) _{+} &=&\sum_{k=0}^{9}\left( \sum_{j=0}^{5}\Lambda
_{j,k-j}\right) D^{k}  \notag \\
&=&\sum_{k=0}^{9}\beta _{k}D^{k}
\end{eqnarray}%
where \ \ \ \
\begin{equation}
\Lambda _{j,k}=\left\{
\begin{array}{cc}
0 & \text{ if }j>5\text{ or }k<0\text{ or }k>4 \\
\neq \text{ }0 & \text{otherwise}%
\end{array}%
\right.
\end{equation}%
The non vanishing values of the superfields $\Lambda _{j,k}$ are
\begin{eqnarray}
\ \Lambda _{0,i} &=&\lambda _{i}^{(5)}+J\lambda _{i}^{^{\prime \prime \prime
}}+Q\lambda _{i}^{^{\prime \prime }}+T\lambda _{i}^{^{\prime }}+W\lambda _{i}
\notag \\
\Lambda _{1,i} &=&(-)^{i}\left[ \lambda _{i}^{(4)}+J\lambda _{i}^{^{\prime
\prime }}+T\lambda _{i}\right]   \notag \\
\Lambda _{2,i} &=&2\lambda _{i}^{^{\prime \prime \prime }}+J\lambda
_{i}^{^{\prime }}+Q\lambda _{i} \\
\Lambda _{3,i} &=&(-)^{i}\left[ 2\lambda _{i}^{^{\prime \prime }}+J\lambda
_{i}\right]   \notag \\
\Lambda _{4,i} &=&\lambda _{i}^{^{\prime }}  \notag \\
\Lambda _{5,i} &=&(-)^{i}\lambda _{i}  \notag
\end{eqnarray}%
Therefore we have
\begin{eqnarray}
\beta _{0} &=&\Lambda _{0,0}  \notag \\
\beta _{1} &=&\Lambda _{0,1}+\Lambda _{1,0}  \notag \\
\beta _{2} &=&\Lambda _{0,2}+\Lambda _{1,1}+\Lambda _{2,0}  \notag \\
\beta _{3} &=&\Lambda _{0,3}+\Lambda _{1,2}+\Lambda _{2,1}+\Lambda _{3,0}
\notag \\
\beta _{4} &=&\Lambda _{0,4}+\Lambda _{1,3}+\Lambda _{2,2}+\Lambda
_{3,1}+\Lambda _{4,0} \\
\beta _{5} &=&\Lambda _{1,4}+\Lambda _{2,3}+\Lambda _{3,2}+\Lambda
_{4,1}+\Lambda _{5,0}  \notag \\
\beta _{6} &=&\Lambda _{2,4}+\Lambda _{3,3}+\Lambda _{4,2}+\Lambda _{5,1}
\notag \\
\beta _{7} &=&\Lambda _{3,4}+\Lambda _{4,3}+\Lambda _{5,2}  \notag \\
\beta _{8} &=&\Lambda _{4,4}+\Lambda _{5,3}  \notag \\
\beta _{9} &=&\Lambda _{5,4}  \notag
\end{eqnarray}%
The differential operator $V_{P}(L)$ then reads
\begin{equation}
V_{P}(L)=\sum_{i=0}^{n}\left( \beta _{k}-\gamma _{k}\right) D^{k}
\end{equation}%
This is easily seen, because \bigskip \bigskip
\begin{equation}
\beta _{9}=\gamma _{9},\text{ }\beta _{8}=\gamma _{8},\text{ }\beta
_{7}=\gamma _{7},\text{ }\beta _{6}=\gamma _{6},\text{ }\beta _{5}=\gamma
_{5}
\end{equation}%
To define an $\ sl\left( 3\mid 3\right) ^{(1)}$ affine structure on $V_{P}(L)
$, one must require the vanishing of the super trace, which is equivalent to
set
\begin{equation}
Sres\left\{ L,P\right\} =0
\end{equation}%
or simply
\begin{equation}
\beta _{4}-\gamma _{4}=0
\end{equation}%
We find
\begin{equation}
X_{5}^{^{\prime }}(\hat{z})=%
\begin{array}[t]{l}
-X_{1}^{(5)}-X_{2}^{(4)}+2\left( X_{3}\right) ^{\prime \prime \prime
}+\left( X_{1}J\right) ^{\prime \prime \prime }+2\left( X_{4}\right)
^{\prime \prime } \\
\\
+\left( X_{2}J\right) ^{\prime \prime }-\left( X_{1}Q\right) ^{\prime \prime
}-\left( X_{1}T\right) ^{\prime }-\left( X_{3}J\right) ^{\prime } \\
\end{array}%
\end{equation}%
or equivalently:
\begin{equation}
X_{5}(\hat{z})=%
\begin{array}[t]{l}
\left( \frac{\delta f}{\delta W}\right) ^{(4)}-\left( \frac{\delta f}{\delta
T}\right) ^{(3)}-2\left( \frac{\delta f}{\delta Q}\right) ^{\prime \prime
}-\left( \frac{\delta f}{\delta W}J\right) ^{\prime \prime }+2\left( \frac{%
\delta f}{\delta J}\right) ^{\prime } \\
\\
+\left( \frac{\delta f}{\delta T}J\right) ^{\prime }+\left( \frac{\delta f}{%
\delta W}Q\right) ^{\prime }+\left( \frac{\delta f}{\delta W}T\right)
+\left( \frac{\delta f}{\delta Q}J\right)  \\
\end{array}%
\end{equation}%
Putting the constraint equation eq(4.56) into the expressions of the non
vanishing values of $\left( \beta _{k}-\gamma _{k}\right) $ one can show
that $V_{P}(L)$ is a differential operator of degrees $\left( 0,3\right) $ :
\begin{equation}
V_{P}(L)=A_{3}D^{3}+A_{2}D^{2}+A_{1}D+A_{0}
\end{equation}%
where $A_{i}=\beta _{i}-\gamma _{i}$ are superfields of dimension$\ \Delta
(A_{i})=\frac{5-i}{2}$ . We give here below the explicit form of the terms $%
A_{i}$ needed in the derivation of the $N=2$ supersymmetric W$_{3}$-algebra.
\begin{equation}
\begin{array}{cc}
A_{3} & =%
\begin{array}[t]{l}
-2X_{1}^{\left( 6\right) }-3X_{2}^{\left( 5\right) }+3X_{3}^{\left( 4\right)
}+2\left( X_{1}J\right) ^{\left( 4\right) }+6X_{4}^{\left( 3\right)
}+3\left( X_{2}J\right) ^{\left( 3\right) } \\
\\
-2\left( X_{1}Q\right) ^{\left( 3\right) }+\left( X_{2}Q\right) ^{\prime
\prime }-\left( X_{3}J\right) ^{\prime \prime }-2\left( X_{1}T\right)
^{\prime \prime }-\left( X_{4}J\right) ^{\prime }-\left( X_{2}T\right)
^{\prime } \\
\\
+JX_{4}^{\prime }+QX_{3}^{\prime }+TX_{2}^{\prime }+WX_{1}^{\prime
}+2QX_{4}+2WX_{2} \\
\end{array}%
\end{array}%
\end{equation}%
\begin{equation}
\begin{array}{cc}
A_{2} & =%
\begin{array}[t]{l}
-X_{2}^{\left( 6\right) }+3X_{4}^{\left( 4\right) }+\left( X_{2}J\right)
^{\left( 4\right) }+\left( X_{2}Q\right) ^{\left( 3\right) }+JX_{4}^{\prime
\prime }+TX_{2}^{\prime \prime }-WX_{1}^{\prime \prime } \\
\\
-\left( X_{3}Q\right) ^{\prime \prime }-\left( X_{4}Q\right) ^{\prime
}-\left( X_{2}W\right) ^{\prime }+\left( X_{2}T\right) ^{\prime \prime
}-2\left( X_{1}W\right) ^{\prime \prime }-QX_{3}^{\prime \prime } \\
\end{array}%
\end{array}%
\end{equation}%
\begin{equation}
\begin{array}{cc}
A_{1} & =%
\begin{array}[t]{l}
-X_{1}^{\left( 8\right) }-2X_{2}^{\left( 7\right) }+X_{3}^{\left( 6\right)
}-JX_{1}^{\left( 6\right) }+\left( X_{1}J\right) ^{\left( 6\right)
}+3X_{4}^{\left( 5\right) }-2JX_{2}^{\left( 5\right) } \\
\\
-\left( X_{1}Q\right) ^{\left( 5\right) }-QX_{1}^{\left( 5\right) }+2\left(
X_{2}J\right) ^{\left( 5\right) }+\left( X_{2}Q\right) ^{\left( 4\right)
}+JX_{3}^{\left( 4\right) }-QX_{2}^{\left( 4\right) } \\
\\
+J\left( X_{1}J\right) ^{\left( 4\right) }-\left( X_{1}T\right) ^{\left(
4\right) }+2J\left( X_{2}J\right) ^{\left( 3\right) }+3JX_{4}^{\left(
3\right) }+Q\left( X_{1}J\right) ^{\left( 3\right) } \\
\\
+QX_{3}^{\left( 3\right) }-TX_{2}^{\left( 3\right) }-J\left( X_{1}Q\right)
^{\left( 3\right) }-WX_{1}^{\left( 3\right) }-\left( X_{2}T\right) ^{\left(
3\right) }+J\left( X_{2}Q\right) ^{\prime \prime } \\
\\
-Q\left( X_{1}Q\right) ^{\prime \prime }+Q\left( X_{2}J\right) ^{\prime
\prime }+QX_{4}^{\prime \prime }-WX_{2}^{\prime \prime }-J\left(
X_{1}T\right) ^{\prime \prime }-\left( X_{3}T\right) ^{\prime \prime
}-X_{3}^{\prime \prime }T \\
\\
-\left( X_{2}W\right) ^{\prime \prime }+T\left( X_{2}J\right) ^{\prime
}+TX_{4}^{\prime }+W\left( X_{1}J\right) ^{\prime }-J\left( X_{2}T\right)
^{\prime }+WX_{3}^{\prime } \\
\\
-\left( X_{4}T\right) ^{\prime }-Q\left( X_{1}T\right) ^{\prime }+2T\left(
X_{2}Q\right) -2WX_{1}Q+2WX_{2}R+2WX_{4} \\
\end{array}%
\end{array}%
\end{equation}%
\begin{equation}
\begin{array}{cc}
A_{0} & =%
\begin{array}[t]{l}
-X_{2}^{\left( 8\right) }+2X_{4}^{\left( 6\right) }+\left( X_{2}R\right)
^{\left( 6\right) }-RX_{2}^{\left( 6\right) }+\left( X_{2}Q\right) ^{\left(
5\right) }-QX_{2}^{\left( 5\right) }+2RX_{4}^{\left( 4\right) } \\
\\
+R\left( X_{2}R\right) ^{\left( 4\right) }-TX_{2}^{\left( 4\right) }-\left(
X_{1}W\right) ^{\left( 4\right) }+X_{1}^{\left( 4\right) }W+R\left(
X_{2}Q\right) ^{\left( 3\right) }+2QX_{4}^{\left( 3\right) } \\
\\
-\left( X_{2}W\right) ^{\left( 3\right) }+Q\left( X_{2}R\right) ^{\left(
3\right) }+Q\left( X_{2}Q\right) ^{\prime \prime }+2TX_{4}^{\prime \prime
}-\left( X_{3}W\right) ^{\prime \prime }-2X_{3}^{\prime \prime }W \\
\\
+T\left( X_{2}R\right) ^{\prime \prime }-R\left( X_{1}W\right) ^{\prime
\prime }-\left( X_{1}R\right) ^{\prime \prime }W-\left( X_{4}W\right)
^{\prime }-R\left( X_{2}W\right) ^{\prime } \\
\\
-Q\left( X_{1}W\right) ^{\prime }+\left( X_{1}Q\right) ^{\prime }W+T\left(
X_{2}Q\right) ^{\prime }+2WX_{2}Q \\
\end{array}%
\end{array}%
\end{equation}%
An important step towards deriving the supersymmetric Gelfand Dickey Poisson
brackets associated to the affine Lie super algebra $sl\left( 3\mid 3\right)
^{(1)}$ generated by the superfields $\left( J,Q,T,W\right) \equiv \left(
U_{1},U_{\frac{3}{2}},U_{2},U_{\frac{5}{2}}\right) $ is towards the
superconformal transformation of the super Lax operator $L^{(0,5)}$ namely:
\begin{equation}
\begin{array}{lll}
\tilde{L} & = & \tilde{D}^{5}+\tilde{J}\tilde{D}^{3}+\tilde{Q}\tilde{D}^{2}+%
\tilde{T}\tilde{D}+\tilde{W} \\
& = & \left( D\tilde{\theta}\right) ^{-3}\left[ D^{5}+JD^{3}+QD^{2}+TD+W%
\right] \left( D\tilde{\theta}\right) ^{-2}%
\end{array}%
\end{equation}%
Identifying both sides of this equation, one obtain:
\begin{equation}
\begin{array}{lll}
J & = & \left( D\tilde{\theta}\right) ^{2}\tilde{J} \\
&  &  \\
Q & = & \left( D\tilde{\theta}\right) ^{3}\tilde{Q}+\tilde{J}\left( D\tilde{%
\theta}\right) \left( D^{2}\tilde{\theta}\right) +3S\left( z,\tilde{z}%
\right)  \\
&  &  \\
T & = &
\begin{array}[t]{l}
\left( D\tilde{\theta}\right) ^{4}\tilde{T}-\left( D\tilde{\theta}\right)
^{2}\left( D^{2}\tilde{\theta}\right) \tilde{Q}+\left( D\tilde{\theta}%
\right) \left( D^{3}\tilde{\theta}\right) \tilde{J} \\
+\left( DS\left( z,\tilde{z}\right) \right)
\end{array}
\\
&  &  \\
W & = &
\begin{array}[t]{l}
\tilde{W}\left( D\tilde{\theta}\right) ^{5}+2\tilde{T}\left( D\tilde{\theta}%
\right) ^{3}\left( D^{2}\tilde{\theta}\right) +\left( D\tilde{\theta}\right)
^{2}\left( D^{3}\tilde{\theta}\right) \tilde{Q} \\
+2\tilde{J}\left[ \left( D\tilde{\theta}\right) \left( D^{4}\tilde{\theta}%
\right) -D^{2}\tilde{\theta}D^{3}\tilde{\theta}\right] +2\left( D^{2}S\left(
z,\tilde{z}\right) \right)
\end{array}%
\end{array}%
\end{equation}%
The $N=2$ super-$W_{3}$ algebra is generated by the superfields $\left(
J,Q,T,W\right) \equiv \left( U_{1},U_{\frac{3}{2}},U_{2},U_{\frac{5}{2}%
}\right) $ and gives rise to the supermultiplet $\left( 1,\left( \frac{3}{2}%
\right) ^{2},2,\left( \frac{5}{2}\right) ^{2},3\right) $ The computation of
the super GD bracket generating the $N=2$ version of the $w_{3}$-algebra
contains long and complicated expressions that we are not putting in the
manuscript. However, we note that we have present a consistent algebraic
analysis and several important properties as well as the crucial steps
necessary to derive the super GD bracket of the $N=2$ super $w_{3}$-algebra.
The principal key in this context is given by the hamiltonian operator $%
V_{P}(L)$ that we have compute completely. \bigskip

\section{Concluding Remarks}

We use in this work a consistent and systematic analysis developed in
previous occasions \cite{9} to study an important problem namely the
supersymmetric version of N=2 Gelfand Dickey algebra. The conformal algebra
and its supersymmetric extensions have played a central role in the study of
string dynamics, statistical models of critical phenomena, and more
generally in two dimensional conformal field theories (CFT)\cite{2,3}.
\\\\
These are symmetries generated by conformal spin $s$ currents with
$s\leq 2$. The extension {\`{a}} la Zamolodchikov incorporates
also currents of higher conformal spin $3$ and $5/2$ and gives
then the super $w$-algebra involving besides the usual spin-2
energy momentum tensor, a conformal spin-3 conserved
current\cite{10,11}. The $w$-symmetry, which initially was
identified as the symmetry of the critical three states Potts
model, has also been realized as the gauge symmetry of the
so-called $w_{3}$ gravity. In relation to integrable systems,
these symmetries are shown to play a pioneering role as their
existence gives, in some sense, a guarantee of integrability.
\\\\
All these physical ideas are behind our initiative to renew the
interest in the supersymmetric version of the Gelfand-Dickey
algebra although the underlying computations are very tedious and
complicated. We focuss in nearest occasion to go beyond these
extensions and apply our analysis to integrable systems in non
trivial symmetries and geometries.


\bigskip
\newpage
\begin{thebibliography}{99}
\bibitem{1} For reviews see for instance: \newline
L. D. Faddeev, L. A. Takhtajan, \emph{Hamiltonian methods and the theory of
solitons}, Springer (1987), A. Das, I\emph{ntegrable Models}, World
scientific 1989.

\bibitem{2} A. A. Belavin, A. M. Polyakov, A. B. Zamolodchikov, Nucl Phys.
B241(1984);\newline
V. S. Dotsenko, V. A. Fateev, Nucl Phys. B240 [FS12], 312-348 (1984);

\bibitem{3} P. Ginsparg, Applied Conformal field Theory, Les houches
Lectures(1988).

\bibitem{4} B. A. Kupershmidt, Phys. Lett. A102, 213(1984);\newline
Y. I. Manin, A. O. Radul, Comm. Math. Phys.98, 65 (1985).

\bibitem{5} P. Mathieu, J. Math. Phys. 29, 2499(1988);\newline
W. Oevel, Z. Popowicz, Comm. math. Phys. 139, 441(1991).

\bibitem{6} E. Ivanov, S. Krivonos, Lett. Math. Phys. 7 (1983) 523; 8 (1984)
345,\newline
E.Ivanov, S.Krivonos, V. Leviant, J. Phys. A: Math. Gen. 22 (1989) 345,
\newline
E. Ivanov, S. Krivonos, V. Leviant, J. Phys. A: Math. Gen. 22 (1989) 4201,
\newline
K. Kobayashi, T. Uematsu, Phys. Lett. B275 (1992) 361,\newline
H. Aratyn, L.A. Ferreira, J.F. Gomes, A.H. Zimerman, Phys. Lett. B281
(1992)245,\newline
E. Ivanov, F. Toppan, 
hep-th/9303073, Phys.Lett. B309 (1993) 289-296

\bibitem{7} F. Delduc, E. Ivanov 
hep-th/9301024, Phys.Lett. B309 (1993) 312-319, \newline
E.Ivanov, S. Krivonos, A.Sorin Mod. Phys. Lett. A10 (1995) 2439,
hep-th/9505142\newline
V. Derjagin, A. Leznov, A. Sorin,
Preprint JINR E2-96-410, hep-th/9611108,\newline
E. Ivanov and S. Krivonos,
hep-th/9609191, Phys.Lett. A231 (1997) 75-81, \newline
F. Delduc, E. Ivanov, S. Krivonos
J.Math.Phys. 37 (1996) 1356-1381; 38 (1997) 1224

\bibitem{8} E. H. Saidi, M. B. Sedra, Class. Quant. Grav.10, 1937-1946(1993);%
\newline
E. H. Saidi, M. B. Sedra, Int. Jour. Mod. Phys. A9, 891-913(1994).

\bibitem{9} E. H. Saidi, M. B. Sedra, J. Math. Phys. 35, 3190(1994);\newline
M. B. Sedra, J. Math. Phys. 37, 3483(1996).

\bibitem{10} A. B. Zamolodchikov, Teo. Math. Fiz.65, 374(1985);\newline
V. A. Fateev, S. Lukyanov, Int. Jour. Mod. Phys. A 3, 507(1988).

\bibitem{11} P. Bouwknegt, K. Schoutens, Phys. Rep. \textbf{223}. 183(1993)
and references therein.

\bibitem{12} J. L. Gervais, Phys. lett. B 160, 277(1985),\newline
A. Bilal and J. L. Gervais, Phys. lett. B 206, 412(1988); \newline
A. Bilal and J. L. Gervais, Nucl. Phys. B 314, 597(1989); \newline
I. Bakas, Nucl. Phys. B 302, 189(1988).

\bibitem{13} B. Khesin, I. Zakharevich, Commun. Math. Phys. 171, 475 (1995)%
\newline
B. Khesin, I. Zakharevich, hep-th/9311125,\newline
P. I. Etingof, B. A. Khesin, arXiv:hep-th/9312123.

\bibitem{14} P. Mathieu, Phys. lett. B 208, 101(1988).

\bibitem{15} K. Yamagishi, Phys. lett. B 259, 436(1991),\newline
F. Yu and Y. S Wu, Phys. lett. B 236, 220(1991),\newline
A. Das, W. J. Huang, S. Panda, Phys. lett. B 271, 109(1991),\newline
A. Das, E. Sezgin, S. J. Sin, Phys. lett. B 277, 435(1992),\newline
I. Bakas, B. Khesin, E. Kiritsis, Commun. Math. Phys. 151, 233 (1993).

\bibitem{16} I. M. Gelfand, V. Sokolov, J. Sov. Math.30, 1975(1985),\newline
I. M. Gelfand, V. Sokolov, Funkt. Anal. Priloz.10, 13(1976),\newline
I. M. Gelfand, V. Sokolov, Funkt. Anal. Priloz. 13, 13(1979).

\bibitem{17} P. Di-Francessco, C. Itzykson, J.B. Zuber, Comm. Math. Phys.
140, 543(1991).

\bibitem{18} K. Huitu, D. Nemeschansky, Mod. Phys. Lett. A6 (1991) 3179-3190.

\bibitem{19} T. Inami, H. Kanno, Commun. Math. Phys. 136 (1991) 519; \newline
T. Inami, H. Kanno, Nucl. Phys. B359 (1991) 201.

\bibitem{20} P. Grozman, D. Leites, I. Shchepochkina, Acta Math. Vietnamica
26, 27 (2005).

\bibitem{21} D. Leites, Theor. Math. Phys. 52, 764 (1982) [Teor. Mat. Fiz.
52, 225 (1982)],\newline
P. Grozman, D. Leites, Czech. J. Phys. 51, 1 (2001) [arXiv:hep-th/9702073].

\bibitem{22} A.N. Leznov, M.V. Saveliev, Commun. Math. Phys. 74 (1980) 11,%
\newline
A.N. Leznov, M.V. Saveliev, Lett. Math.Phys. 3 (1979) 489,\newline
P. Mansfield, Nucl. Phys. B208 (1982) 277,\newline
P. Mansfield, Nucl. Phys. B222 (1983) 419,\newline
D. Olive, N. Turok, Nucl. Phys. B257 [FS14] (1986) 277,\newline
D. Olive, N. Turok, Nucl. Phys. B265 [FS15](1986) 469,

\bibitem{23} A.B. Bilal, J.L. Gervais, Nucl. Phys. B318 (1989) 579,\newline
O. Babelon, Phys. Lett. B215 (1988) 523

\bibitem{24} T. Eguchi, S-K. Yang, Phys. Lett. B224 (1989) 373,\newline
T.J. Hollowood, P. Mansfield, Phys. Lett. B226 (1989) 73

\bibitem{25} V. G. Drinfeld, V. V. Sokolov, Sov. J. Math. 38, 1975 (1985),

\bibitem{26} J. Evans and T. Hollowood, Nucl. Phys. B 352, 529 (1991),%
\newline
H. Noharan, K. Mohri, ibid. 349, 529 (1991).

\bibitem{27} E.~H.~Saidi, M.~B.~Sedra, Mod. Phys. Lett. A 9, 3163 (1994),
hep-th/0512220.

\bibitem{28} E.~H.~Saidi, M.~B.~Sedra, J.~Zerouaoui,
Class.\ Quant.\ Grav.\ \textbf{12}, 2705 (1995);%
\newline
E.~H.~Saidi, M.~B.~Sedra, J.~Zerouaoui,
Class.\ Quant.\ Grav.\ \textbf{12} (1995) 1567.

\bibitem{29} M.~B.~Sedra,
Nucl.\ Phys.\ B \textbf{740}, 243 (2006), hep-th/0508236.
\newline
O.~Dafounansou, A.~El Boukili, M.~B.~Sedra,
Chin.\ J.\ Phys.\ \textbf{44}, 274 (2006), 
\newline
A.~El Boukili, E.~H.~Saidi and M.~B.~Sedra,
hep-th/0610123. 
\newline
A.~Boulahoual and M.~B.~Sedra, 
Afr.\ J.\ Math.\ Phys.\ \textbf{2} (2005) 111. 
\newline
A.~Boulahoual and M.~B.~Sedra,
J.\ Math.\ Phys.\ \textbf{44}, 5888 (2003), hep-th/0308079.
\newline
A.~Boulahoual and M.~B.~Sedra,
Chin.\ J.\ Phys.\ \textbf{43}, 408 (2005) hep-th/0208200.
\newline
A.~Boulahoual and M.~B.~Sedra,
Chin.\ J.\ Phys.\ \textbf{42}, 591 (2004) hep-th/0104086.
\end{thebibliography}
\end{document}